\documentclass[preprint]{iucr}              

     \journalcode{J}              

\usepackage{graphicx}
\usepackage{amssymb}
\usepackage{amsmath}
\usepackage[utf8]{inputenc}
\usepackage{url}
\usepackage{dcolumn}
\newcolumntype{.}{D{.}{.}{-1}}
\graphicspath{{./figures/}}

\newcommand{\iAA}{\AA\ensuremath{^{-1}}}
\newcommand{\Qmaxinst}{\ensuremath{Q_\mathit{maxinst}}}
\newcommand{\Qdamp}{\ensuremath{Q_\mathit{damp}}}
\newcommand{\rpoly}{\ensuremath{r_\mathit{poly}}}
\newcommand{\dothead}[1]{\multicolumn{1}{r}{#1}}

\begin{document}                  

\title{PDFgetX3: A rapid and highly automatable program for processing powder diffraction data into total scattering pair distribution functions}

\author[a]{P.}{Juh\'as}
\author[a]{T.}{Davis}
\author[a]{C.~L.}{Farrow}
\cauthor[a,b]{S.~J.~L.}{Billinge}{sb2896@columbia.edu}{}

\aff[a]{Department of Applied Physics and Applied Mathematics, Columbia University, New York, New York, 10027, USA}
\aff[b]{Condensed Matter Physics and Materials Science Department, Brookhaven National Laboratory, Upton, New York, 11973, USA}

\date{\today}
\maketitle


\begin{abstract}
PDFgetX3 is a new software application for converting X-ray powder
diffraction data to atomic pair distribution function (PDF).  PDFgetX3 has
been designed for ease of use, speed and automated operation. 
The software can readily process hundreds of X-ray patterns within few
seconds and is thus useful for high-throughput PDF studies, that measure
numerous datasets as a function of time, temperature or
other environment parameters.  In comparison to the preceding
programs, PDFgetX3 requires fewer inputs, less user experience
and can be readily adopted by novice users.  The live-plotting
interactive feature allows to assess the effects of calculation
parameters and select their optimum values.  PDFgetX3
uses an \textit{ad-hoc} data correction method, where the
slowly-changing structure independent signal is filtered out
to obtain coherent X-ray intensities that contain structure
information.  The outputs from PDFgetX3 have been verified by
processing experimental PDFs from inorganic, organic and nanosized
samples and comparing them to their counterparts from previous established
software.  In spite of different algorithm, the obtained PDFs were
nearly identical and yielded highly similar results when used in structure
refinement.  PDFgetX3 is written in Python language and features
well documented, reusable codebase.  The software can be used
either as standalone application or as a library of PDF-processing
functions that can be called on from other Python scripts.  The software
is free for open academic research, but requires paid license for
commercial use.
\end{abstract}



\section{Introduction}
\label{Introduction}

With the increased interest in producing and exploiting nanostructured materials, it is necessary to expand the methods that go beyond crystallography \cite{billi;p10} for characterizing their atomic scale structure.  In recent years, total scattering and atomic pair distribution function (PDF) analysis \cite{egami;b;utbp12} has emerged as a popular and powerful tool for this purpose \cite{billi;cc04,billi;jssc08,young;jmc11}.  To satisfy this demand a number of X-ray and neutron beamlines
dedicated to
or optimized for, such measurements have emerged \cite{egami;b;utbp12}, and manufacturers of laboratory X-ray sources are also beginning to market instruments for this kind of measurement.  Especially with the use of 2D detectors, modern beamlines are yielding total scattering data at unprecedented rates allowing detailed parametric and time-resolved total scattering studies to be carried out in special environments
\cite{chupa;jacs04,chupa;jac07,jense;jacs12,redmo;ecssl12}.
A bottleneck in further growth of the method is now the lack of robust and automatable software for creating PDFs from the raw data, currently a computationally and user-intensive process \cite{egami;b;utbp12}.

This can be illustrated by considering one of the most widely used software programs for this purpose, PDFgetX2 \cite{qiu;jac04i}. The program offers users a great deal of flexibility and control in choosing exactly which corrections to apply to X-ray scattering intensities in order to convert them to PDFs. However, due to the myriad of options available to users as well as the esoteric nature of many of the corrections \cite{egami;b;utbp12}, PDF generation requires considerable user input and expertise in arcane details of the technique. Although the software has a graphical user interface, it is a time intensive process to carry out the corrections, with many possibilities for input errors, and the process can't be easily automated for high throughput of many data sets.

In this paper, we describe a new software program, PDFgetX3, which implements an {\it ad-hoc} data reduction algorithm \cite{billi;aca12} that requires little user input, generates PDFs in a fraction of a second, and can be straightforwardly automated to batch-process thousands of PDFs.  Here we show that in the physically relevant region of the PDF it produces quantitatively accurate PDFs that are the same as those obtained using PDFgetX2 for the cases shown, and which yield refined structural parameters that are also indistinguishable from those refined from PDFgetX2 determined PDFs.



The intensities measured in a total scattering experiment, $I_{m}(Q)$, can be expressed as \cite{billi;aca12}
\begin{equation}
\label{eq;Iexp}
I_{m}(Q) = a(Q) I_{c}(Q) + b(Q),
\end{equation}
where $I_{c}(Q)$ is the coherent scattering intensity, which contains all of the structural information about the sample, and $a(Q)$ and $b(Q)$ are multiplicative and additive corrections to the measured intensity, which do not contain structural information \cite{billi;aca12}. Examples of the additive contributions are incoherent Compton scattering and background scattering from the sample container. Examples of the multiplicative contributions are sample self-absorption and polarization of the X-ray beam. The approach used by PDFgetX2, and other PDF data analysis programs, is to apply known corrections to the $I_{m}(Q)$ to obtain the coherent scattering, $I_{c}(Q)$, which is transformed into the structure function, $S(Q)$ according to
\begin{equation}
\label{eq,sq}
S(Q) = \frac{I_{c}(Q) - \langle f(Q)^2\rangle + \langle f(Q)\rangle^2}{\langle f(Q)\rangle^2}.
\end{equation}
Here $f(Q)$ is the atomic scattering factor and the angle brackets indicate an average over all the atom types in the sample.  For the neutron case the $f$'s are replaced by coherent neutron scattering lengths, $b$, in this equation.

The $S(Q)$ is Fourier transformed into the PDF, $G(r)$, according to \cite{farro;aca09}
\begin{equation} \label{eq,gr}
\begin{split}
    G(r) &= \frac{2}{\pi}\int_{Q_{min}}^{Q_{max}} Q[S(Q)-1]\sin Qr\>dQ\\
         &= \frac{2}{\pi}\int_{Q_{min}}^{Q_{max}} F(Q) \sin Qr\>dQ,
\end{split}
\end{equation}
where the quantity $F(Q)=Q[S(Q)-1]$ is  the reduced structure function \cite{warre;b;xd90}.  The many corrections required are discussed in detail in Chapter 5 of Egami and Billinge \cite{egami;b;utbp12}, including
background subtraction, polarization, self-absorption, multiple scattering and Compton scattering, among many others, and these are implemented in PDFgetX2 \cite{qiu;jac04i} and other similar programs
\cite{petko;jac89,petko;jac98,jeong;jac01,soper;jac11}.

It has recently been pointed out \cite{billi;aca12} that sufficient information is known about the general behavior of the correction terms in Eq.~\ref{eq;Iexp}, and about the asymptotic behavior of the resulting $F(Q)$ function, that it may be possible to determine $a(Q)$ and $b(Q)$ through an {\it ad-hoc} approach where they are parameterized and the parameters varied using a regression method in such a way as to yield an accurate $F(Q)$ function.  Here we describe an algorithm for doing this, as well as a software implementation, PDFgetX3, and we show that, indeed, it yields PDFs that are not significantly different from those obtained using PDFgetX2.  The program is fast, easy to use, and highly automatable.


The method was developed initially for analyzing rapid acquisition PDF data
from 2D detectors, though we show below that it is not limited to this application.
It is assumed that the 2D data have been correctly azimuthally
integrated, and multiple frames summed or averaged, to obtain a one-dimensional
intensity vs.\ $Q$ or intensity vs.\ $2\theta$ diffraction. A number of integration
programs exist for this purpose, for example Fit2D \cite{hamme;hpr96}.

The algorithm \cite{billi;p;mopxddacnaac11} starts with raw intensity data measured versus scattering
angle $2\theta$.  At first, the angle is converted to scattering vector
$Q$ and the data are re-sampled to an equidistant $Q$-grid, which is
suitable for a fast Fourier transformation at a later step and
also ensures constant weights in a $Q$-dependent fitting.  Note that resampling
introduces error correlations between points which can be minimized if the data
are azimuthally integrated from 2D directly onto a constant-$Q$ grid \cite{yang;unpub12}.
The background intensities from an empty container are
then re-sampled to the same Q-grid and subtracted from the sample data.
This yields raw intensities from the specimen only; however, which are not
normalized per incident intensity nor per the number of scatterers.
The structure function $S(Q)$ should oscillate around and then approach
unity as $Q$ tends to infinity, which in practice is about $Q=25$~\iAA .  This means
that the difference
\begin{equation}
    S(Q) - 1 = \frac{I(Q)}{\langle f \rangle^2} -
        \frac{\langle f^2 \rangle}{\langle f \rangle^2}
\end{equation}
must oscillate around zero and the normalized intensity
$I(Q) / \langle f \rangle^2$ must be close to
the normal scattering factor
$\langle f^2 \rangle / \langle f \rangle^2$ for any $Q$\@.
The raw sample intensities are therefore rescaled by a least-squares
procedure to approach the normal
scattering factor curve.  A physically correct scattering function
$S(Q)$ should also display proper asymptotic behavior as a derived function
$F(Q) = Q (S(Q) - 1)$, which should oscillate around zero and approach
it with increasing $Q$.  The PDFgetX3 algorithm is based on an assumption
that the experimental function
$S_{m}(Q)$
deviates from the correct value by a slowly changing additive factor
$\beta_{S}(Q)$ such that
\begin{equation}
    S_{m}(Q) - 1 = S(Q) - 1 + \beta_{S}(Q).
\end{equation}
The derived function $F_{m}(Q)$ is then
\begin{equation}
    F_{m}(Q) = Q \left[ S(Q) - 1 + \beta_{S}(Q) \right] =
    F(Q) + Q \beta_{S}(Q).
\end{equation}
Because the correct function $F(Q)$ oscillates around zero, the
error term $\beta_{S}(Q)$ produces a slowly changing, $Q$-increasing background
in $F_{m}(Q)$.  The PDFgetX3 algorithm estimates the background
by modeling the $\beta_{S}(Q)$ function as an $n$-th degree polynomial
$P_n(Q)$, which is then fitted as $Q P_n(Q)$ to the $F_{m}(Q)$
function.  The corrected function $F_{c}(Q)$ is afterwards obtained by
subtracting the polynomial fit
\begin{equation}
F_{c}(Q) = F_{m}(Q) - Q P_n(Q).
\end{equation}
The function $F_{c}(Q)$ shows the correct asymptotic behavior with
$F \rightarrow 0$ for large $Q$ values.  Finally, the $F_{c}(Q)$
signal is converted to $G(r)$ using the fast Fourier transformation
as per Eq.~\ref{eq,gr}.

Since the fitted polynomial is an approximation to the actual
error term $\beta_{S}(Q)$, the corrected function $F_{c}(Q)$ still deviates
from the ideal $F$ by
\begin{equation}
    \Delta F(Q) = F_{c}(Q) - F(Q) =  Q \beta_{S}(Q) - Q P_n(Q),
\end{equation}
and the difference introduces an error signal
$\Delta G(r)$ in the obtained PDF.
The function $Q P_n(Q)$ is an $(n + 1)$st-degree polynomial
approximation to the $Q \beta_{S}(Q)$ function on a fit interval
running from zero to \Qmaxinst, therefore we can assume that
the $\Delta F(Q)$ difference has $(n + 1)$ roots that are essentially
equidistant between $Q = 0$~\iAA\ and \Qmaxinst.  The difference
function $\Delta F(Q)$ switches between
positive and negative values at each root, which roughly
corresponds to
oscillations with a half-period of $\Qmaxinst / n$.
Assuming this to be the maximum $Q$ ``frequency'' in the difference
signal $\Delta F(Q)$, the Fourier transformation
would introduce non-physical signal $\Delta G(r)$ extending
up to
\begin{equation}  \label{eq,rpoly}
    \rpoly = \pi n / \Qmaxinst .
\end{equation}
For a typical RAPDF experimental data, the PDFgetX3 program uses an 8-th
degree polynomial correction with \Qmaxinst~= 28~\iAA, which yields
\rpoly~= 0.9~\AA.
Assuming there are no higher frequency aberrations in the data themselves, the error signal $\Delta G(r)$
arising from the polynomial data correction is thus present only
for lengths smaller
than $r \approx 0.9$~\AA, i.e., in a region below the shortest
bond-lengths in most materials.  Furthermore, the polynomial fit cannot accidentally remove real structural
signals from the experimental intensity provided the value of \rpoly\ is chosen to be below
the nearest neighbor bond distance in the material.

Under some experimental conditions such as from lower energy X-ray
sources, or where the experimental take-off angle is very limited,
the instrument $Q$-range \Qmaxinst\ is
much smaller and may increase the error extent
\rpoly\ to physically meaningful distances.  In such cases, the
degree of the correction polynomial $n$ needs to be reduced to
avoid overcorrecting the measured data and
to keep the value of \rpoly\ small.  The PDFgetX3 procedure
uses Eq.~\ref{eq,rpoly}
in reverse, and for a fixed value of the
error extent \rpoly\ and instrument range \Qmaxinst,
it obtains the
degree of correction polynomial as
\begin{equation}
n_r = \rpoly \Qmaxinst / \pi.
\end{equation}
This estimate of the polynomial degree $n_r$ is almost never an integer,
and rounding it to an integer would introduce abrupt changes in the PDF at the
half-integer values.
We would prefer
the PDF to respond smoothly to the \rpoly\ and \Qmaxinst\ parameters.
To simulate a polynomial fit at an arbitrary floating-point degree,
the correction polynomial is therefore refined twice, for an integer
floor and ceiling of $n_r$, and the two fits are then averaged with
the weights given by the distance of $n_r$ from its integer bounds.

\section{Program availability and operation}
\label{Operation}

PDFgetX3 is written in the Python programming language \cite{python;web}. To run, it requires Python 2.5 or later with the NumPy and Matplotlib libraries installed (note that Python 3.0 is not currently supported). It has been tested to work on the Windows, Linux, and Mac operating systems. Information on the installation and operation of the software can be found at the www.diffpy.org website. The command-line version is free for university based researchers conducting open academic research,
but other uses require a paid license.  A version with a graphical user interface (GUI) and online version also under development.  Information can be found at http://diffpy.org.

Because the corrections are {\it ad-hoc}, only minimal information needs to be supplied by the user and this is contained in a configuration file, or can be specified as a command-line argument. In the current implementation the program reads data that are stored in a multi-column text file with the independent variable, $Q$ or $2\theta$ in the first column and the measured intensity in the second.  If the uncertainties on points in the data are known, these may be placed in subsequent columns. The filename for the input file, and a measured background file if one wants to subtract it, must be specified and if the independent variable is $2\theta$ then an X-ray wavelength must also be specified.  The approximate composition of the sample is also specified so that the $f(Q)$ averages may be computed accurately.  A background scale parameter, and $Q_{max}$ to be used in the Fourier transform, are specified, though these have default values and the program works when they are not provided.  The optimal values of some of these parameters may not be known {\it a priori} and the program may be run in an interactive mode where various tuning parameters may be varied by sliding a slider with the resulting PDFs updating in real-time in a plot window.  In this way a user may quickly find the optimal $Q_{max}$ and background scale values that are fed back to the program.  It takes only a few microseconds to complete the corrections on the raw data and so the plots update in real-time as the user adjusts the slider. Some other parameters may also be controlled by the user to obtain the desired output, as described in the manual.

\begin{figure}
    \centering
        \includegraphics[width=0.6\textwidth]{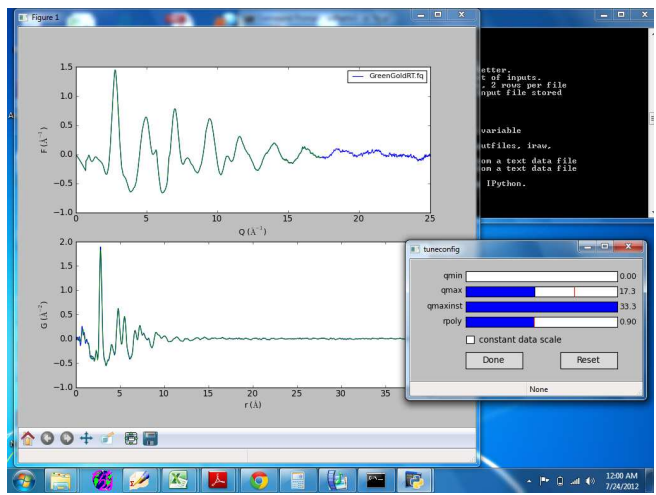}
    \caption{Screenshot of PDFgetX3 in interactive mode. The user selected to plot $F(Q)$ and $G(r)$.  These plots get updated in real-time as the user uses the mouse to move the sliders. There are four sliders in this example for $Q_{min}$, $Q_{max}$, \Qmaxinst\ and \rpoly.  The first two are self explanatory. \Qmaxinst\ varies the range over which the correction polynomial is fit and \rpoly\ places an upper bound on the frequency of information that the {\it ad-hoc} procedure can remove by fitting (see \cite{billi;aca12} for details). If the user wishes to subtract a background signal the background scale will also appear as a slider option.}
    \label{fig;screenshot}
\end{figure}

The program has a powerful Python-based command-line interpreter capability, for example, allowing templates to be used for multiple files that have the same filename-root but which iterate in some way in the name, for example, by run number. This makes the automation of data reduction of many hundreds or thousands of datasets rather straightforward.  The program is also written with a well documented API so that programmers can access the functionality of the engine within home written Python scripts of arbitrary complexity.  A screenshot of the program working in interactive mode is shown in Figure~\ref{fig;screenshot}.

\section{Comparison of PDFgetX3 and PDFgetX2 PDFs}

PDFs have been determined with PDFgetX3 for a number of representative samples and compared with those determined from the same data using PDFgetX2.  The resulting PDFs are compared by plotting them on top of each other.  Where possible, structural models have been refined to both PDFs allowing a direct comparison of fit quality and the values of refined structural parameters from each PDF.  The examples include inorganic materials such as bulk nickel and barium titanate, nanostructured $\gamma$-alumina, and bulk and nanocrystalline cadmium selenide, as well as crystalline and nanostructured phases of the organic pharmaceutical carbamazepine. We choose these very different types of materials to show that PDFgetX3 is a robust program that can handle all sorts of high energy X-ray data.

In all cases, PDFs from both programs are made from the same raw data and, where appropriate, use the same input parameters (i.e., $Q_{max}$, X-ray wavelength, chemical composition, and container background). All data sets except the $\gamma$-Al$_{2}$O$_{3}$ were collected at high-energy synchrotron instruments using the rapid acquisition PDF mode (RAPDF) \cite{chupa;jac03} where data are collected on a 2D detector and azimuthally integrated to obtain 1D datasets, however, the synchrotron is not a requirement.  PDFgetX3 can handle data from lab-based XRPD instruments and synchrotron data collected in point-by-point mode such as from high resolution diffractometers such as ID31 at ESRF. To show this the comparison is done for $\gamma$-Al$_{2}$O$_{3}$ data that were collected with a Panalytical laboratory based silver anode diffractometer.


In general, as we will see in the following examples, we find that the PDFs made by the different programs look somewhat different from one another at $r$ values lower than \rpoly . However, in the physically meaningful range beyond the first nearest neighbor peak the PDFs look almost exactly the same.
In the plots the PDFs obtained by the different methods have been rescaled by a constant such that the nearest neighbor peak is the same height between PDFs on the same plot. The \textit{ad-hoc} approach \cite{billi;aca12} does not result in absolutely normalized data and normalization must be carried out by other methods.  A constant scale offset has been shown not to affect the structural information in the PDF \cite{peter;jac03} when it is modeled with a scale factor variable, since the relative scaling of peaks to one another within the same PDF is preserved.

Models were fit to the PDFs by refining a variety of parameters as appropriate, such as lattice parameters, atomic positions, thermal factors, using the program PDFgui \cite{farro;jpcm07}.  In each case, we compare the $R_w$ value as well as the values of the refined parameters from the PDF obtained using PDFgetX2 and PDFgetX3.

\subsection{Nickel and Barium Titanate}

\begin{figure}
    \begin{center}
    \includegraphics[width=0.6\textwidth]{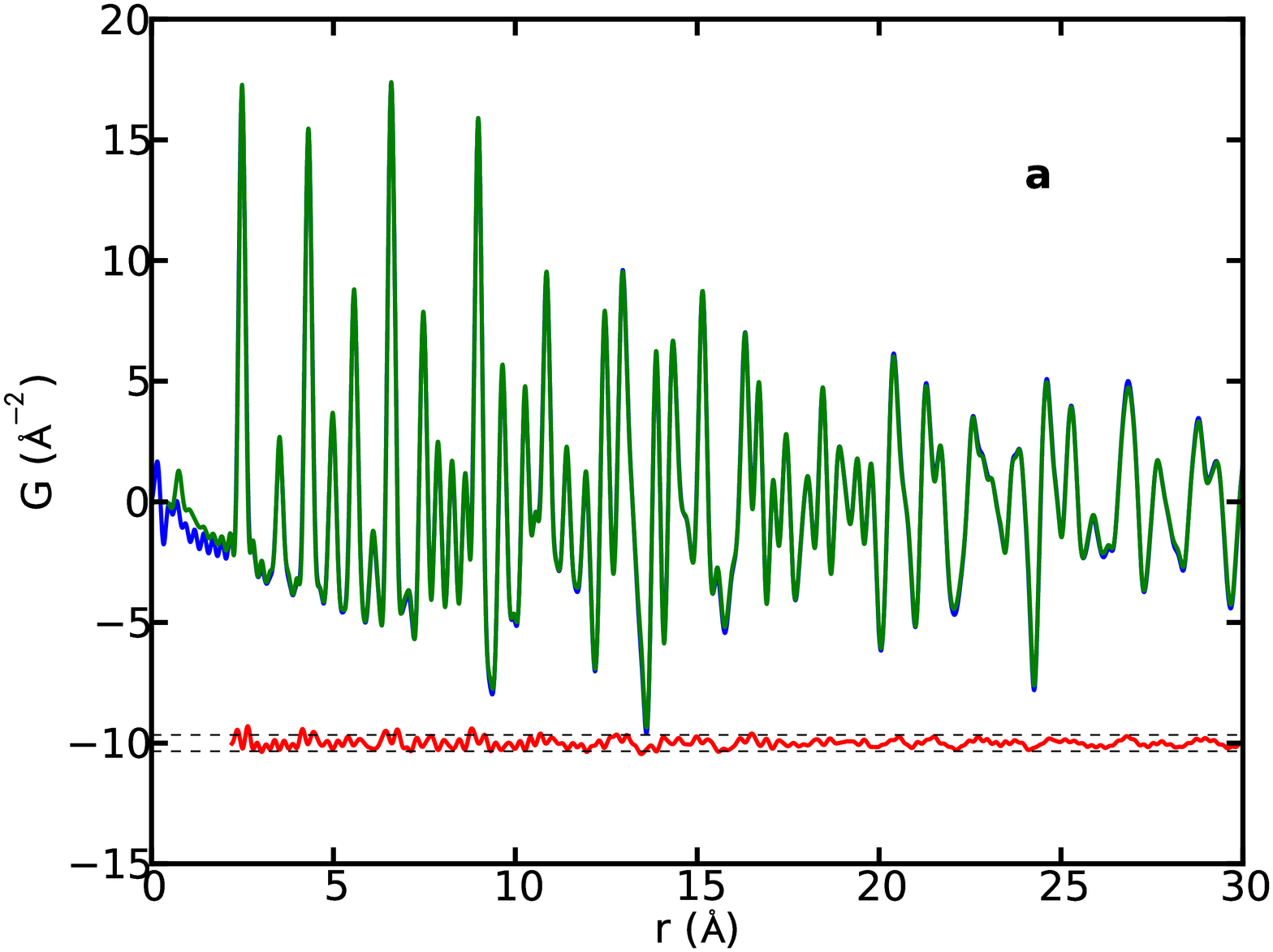}
    \includegraphics[width=0.6\textwidth]{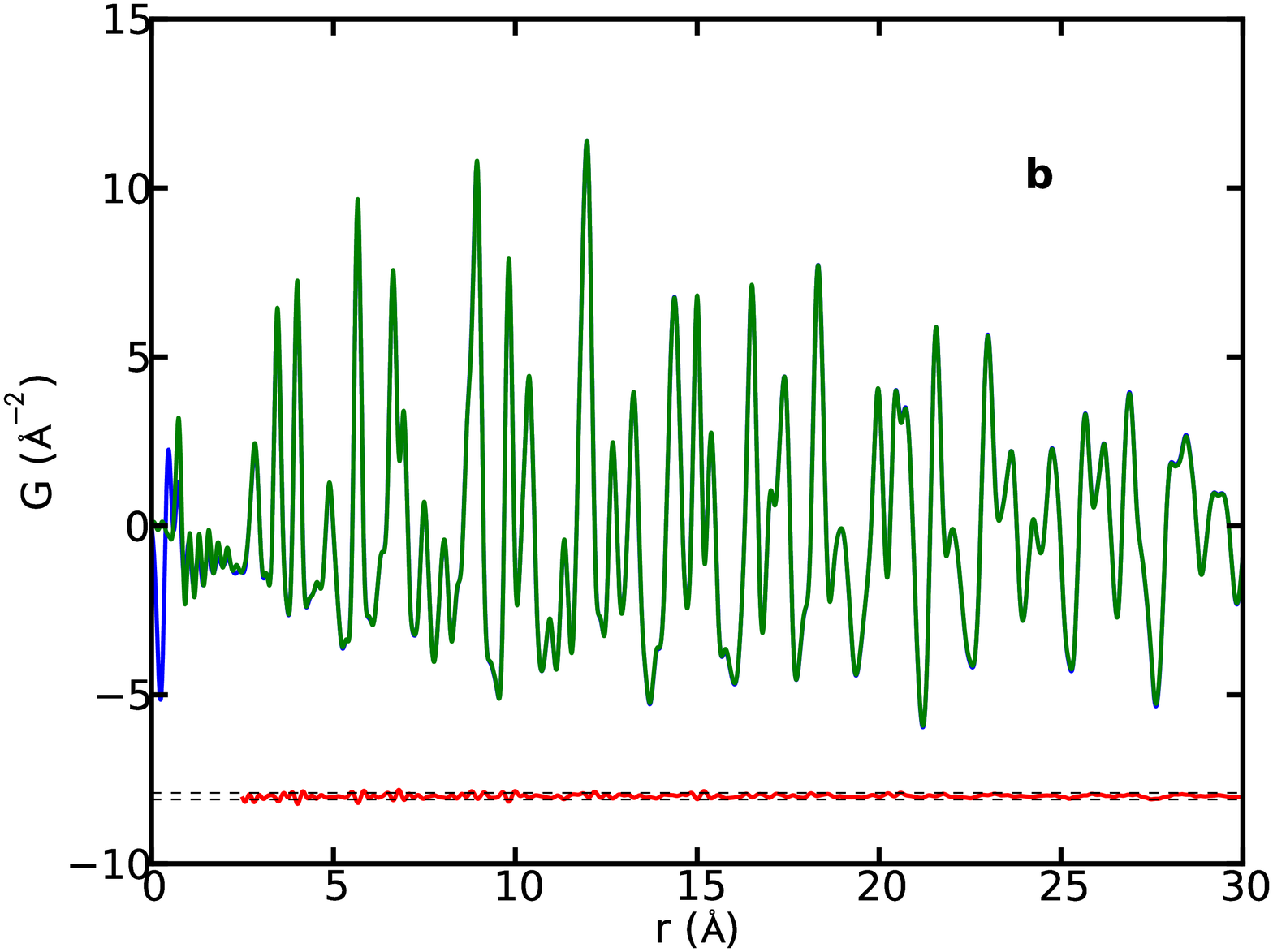}
    \end{center}
    \caption{PDFs of \textbf{(a)} nickel and \textbf{(b)} barium titanate made with PDFgetX2 (blue) and PDFgetX3 (green) with $Q_{max}$ = 26.0~\AA$^{-1}$ in both cases. Difference curve (offset) is in red. The dashed lines represent two standard deviations in the difference curve ($r$ values below the nearest neighbor peaks were not included in the standard deviation calculation).}
    \label{fig;ni-compare}
\end{figure}

First we look at pure nickel (Ni) and barium titanate (BaTiO$_{3}$) in Figure~\ref{fig;ni-compare}.
Both compounds diffract strongly making data corrections less challenging. Figure~\ref{fig;ni-compare}(a) shows the two PDFs of nickel plotted on top of one another. In all the figures the PDF made with PDFgetX2 is in blue and the PDF made with PDFgetX3 is in green.  Here the $Q_{max} = 26.0$~\AA$^{-1}$ in both cases. The difference curve between the two PDFs is plotted offset below in red with dashed lines plotted at $\pm 2\sigma$ as guides to the eye, where $\sigma$ is the standard deviation of the difference computed over the range above \rpoly . We see  only very small differences between the two PDFs after the Ni-Ni nearest neighbor peak (at $r = 2.2$~\AA). We see the same behavior in Figure~\ref{fig;ni-compare}(b) with barium titanate.

The refined parameters from model fits are reproduced in Table~\ref{tab;ni-fit}. In the case of Ni only a few parameters may be varied  because of the simplicity of the structure. Overall, we see very good agreement between most of the parameters and the $R_w$ values.  There are more structural parameters that may be varied in the BaTiO$_3$ case \cite{megaw;ac62} as reproduced in Table~\ref{tab;batio3-fit}. The parameters still agree very well with one another and the quality of the fit as measured by $R_w$ is the same.  We do not report estimated standard deviations on the refined parameters since we do not have reliable error estimates for the data themselves. 
The enhancement of PDFgetX3 to propagate uncertainties on the data and
the problem of extracting reliable uncertainties on integrated powder
data from 2D integrating detectors are being addressed \cite{yang;unpub12},
so we expect this problem to be resolved in the near future.

\begin{table}
\centering
\caption{Comparison of the parameters refined in fitting the Ni model to the PDFs.}
\label{tab;ni-fit}
\begin{tabular}{l . .}
\hline
Parameter  &  \dothead{PDFgetX2}  &  \dothead{PDFgetX3}  \\
\hline
\Qdamp\ (\AA$^{-1}$)         & 0.0554   & 0.0570  \\
$a=b=c$ (\AA)                & 3.5239   & 3.5237  \\
$\delta_{2}$ (\AA$^{2}$)     & 2.52     & 2.71    \\
$U_{iso}$ (\AA$^{2}$)        & 0.00612  & 0.00564 \\
$R_w$                        & 0.0796   & 0.0821  \\
\hline
\end{tabular}
\end{table}

\begin{table}
\centering
\caption{Comparison of the parameters refined in fitting the BaTiO$_{3}$ model to the PDFs.}
\label{tab;batio3-fit}
\begin{tabular}{l . .}
Parameter  &  \dothead{PDFgetX2}  &  \dothead{PDFgetX3}  \\
\hline
\Qdamp\ (\AA$^{-1}$)                & 0.0485   & 0.0491   \\
$a=b$ (\AA)                         & 3.9952   & 3.9952   \\
$c$ (\AA)                           & 4.0399   & 4.0398   \\
$\delta_{2}$ (\AA$^{2}$)            & 4.32     & 4.37   \\
$U_{11,\mathrm{Ba}}$ = $U_{22,\mathrm{Ba}}$ (\AA$^{2}$)
                                    & 0.00516  & 0.00494  \\
$U_{33,\mathrm{Ba}}$ (\AA$^{2}$)    & 0.00454  & 0.00432  \\
$U_{11,\mathrm{Ti}}$ = $U_{22,\mathrm{Ti}}$ (\AA$^{2}$)
                                    & 0.00874  & 0.00839  \\
$U_{33,\mathrm{Ti}}$ (\AA$^{2}$)    & 0.0125   & 0.0121   \\
$U_{11,\mathrm{O}}$ = $U_{22,\mathrm{O}}$ (\AA$^{2}$)
                                    & 0.0113   & 0.0103   \\
$U_{33,\mathrm{O}}$ (\AA$^{2}$)     & 0.0927   & 0.0953   \\
$R_w$                               & 0.118    & 0.121    \\
\hline
\end{tabular}
\end{table}

\subsection{Nanocrystalline $\gamma$-Alumina}

\begin{figure}
    \centering
        \includegraphics[width=\textwidth]{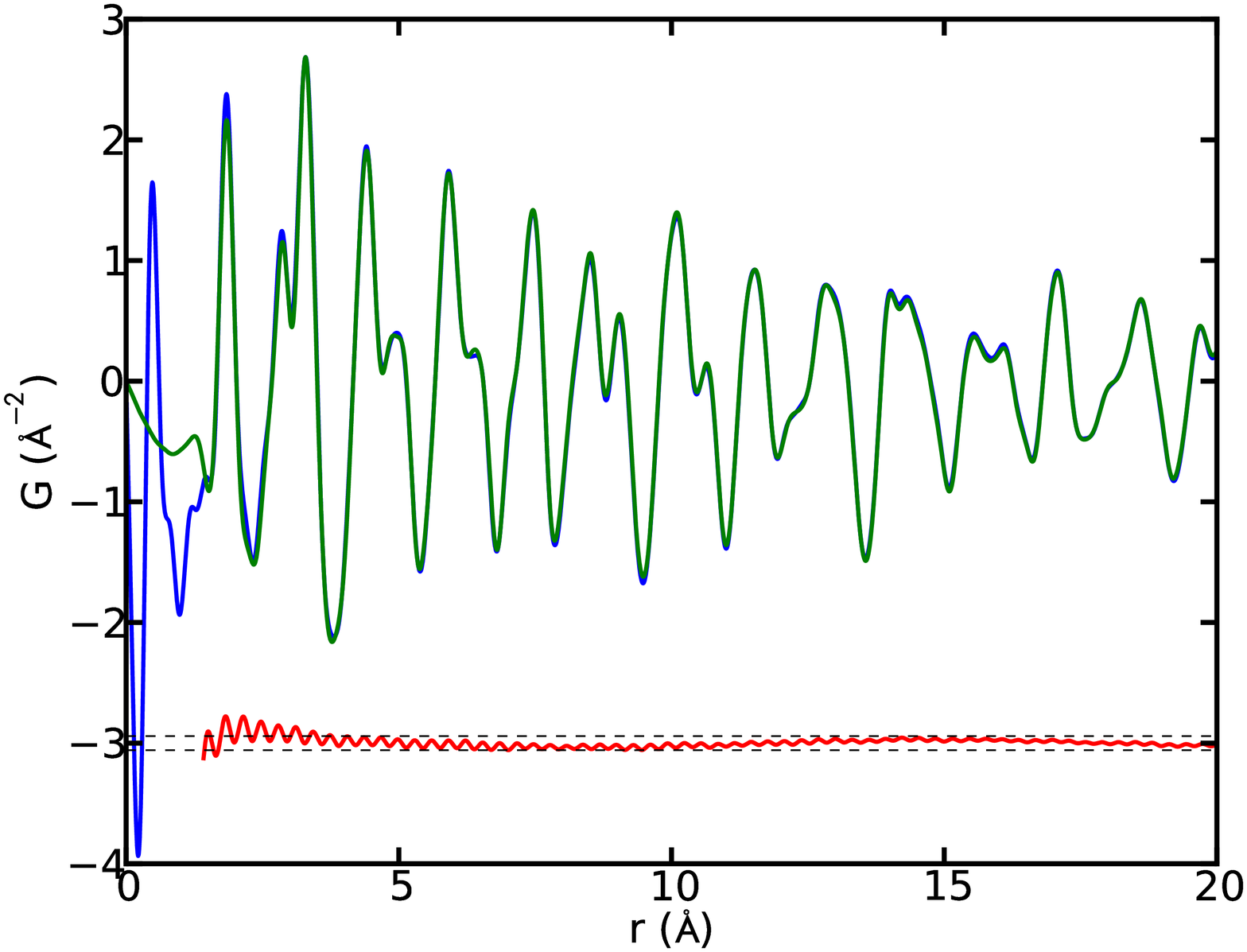}
    \caption{PDFs of $\gamma$-Al$_{2}$O$_{3}$ made with PDFgetX2 (blue) and PDFgetX3 (green) with $Q_{max}$ = 20.5~\AA$^{-1}$ in both cases. Difference curve (offset) is in red. The dashed lines represent two standard deviations in the difference curve ($r$ values below the nearest neighbor peaks were not included in the standard deviation calculation).}
    \label{fig;al2o3-compare}
\end{figure}

Next, we investigate $\gamma$-alumina (Al$_{2}$O$_{3}$) using X-rays from a silver anode diffractometer ($\lambda = 0.56$~\AA). The $\gamma$ phase of Al$_{2}$O$_{3}$ has a local nanocrystalline structure that is different from the structures over longer-range \cite{pagli;cm06ii}. For this reason, a new structure model was developed for the local structure of $\gamma$-Al$_{2}$O$_{3}$ up to $r = 8$~\AA\ (ICSD 173014) \cite{pagli;cm06ii}. Figure~\ref{fig;al2o3-compare} shows the PDFs of $\gamma$-Al$_{2}$O$_{3}$ made with PDFgetX2 and PDFgetX3 over this region. We see very good agreement between the PDFs. In fact, the PDFgetX3 PDF looks better at low $r$ values.

Refined parameters are in Table~\ref{tab;al2o3-fit}. Unlike in previous cases where we tried to use a large $r$ range for our refinement, in this case we refined only over the range $r =1.5 - 8$~\AA\ because the model only applies over this range. For this reason, we wanted to refine few parameters (this is why $U_{iso}$ was used rather than anisotropic thermal factors). Regardless, though, we see very good agreement between the fit results.

\begin{table}
\centering
\caption{Comparison of the parameters refined in fitting the $\gamma$-Al$_{2}$O$_{3}$ model to the PDFs.}
\label{tab;al2o3-fit}
\begin{tabular}{l . .}
Parameter  &  \dothead{PDFgetX2}  &  \dothead{PDFgetX3}  \\
\hline
\Qdamp\ (\AA$^{-1}$)                & 0.0770    & 0.0808    \\
$a$ (\AA)                           & 3.3943    & 3.3941    \\
$b$ (\AA)                           & 2.7796    & 2.7802    \\
$c$ (\AA)                           & 7.0419    & 7.0395    \\
$\delta_{2}$ (\AA$^{2}$)            & 1.13      & 0.991     \\
$U_{iso,\mathrm{O}}$ (\AA$^{2}$)    & 0.0126    & 0.0123    \\
$U_{iso,\mathrm{Al}}$ (\AA$^{2}$)   & 0.0148    & 0.0145    \\
$R_w$                               & 0.164     & 0.166     \\
\hline
\end{tabular}
\end{table}

\subsection{Cadmium Selenide Nanoparticles}

\begin{figure}
    \centering
        \includegraphics[width=0.5\textwidth]{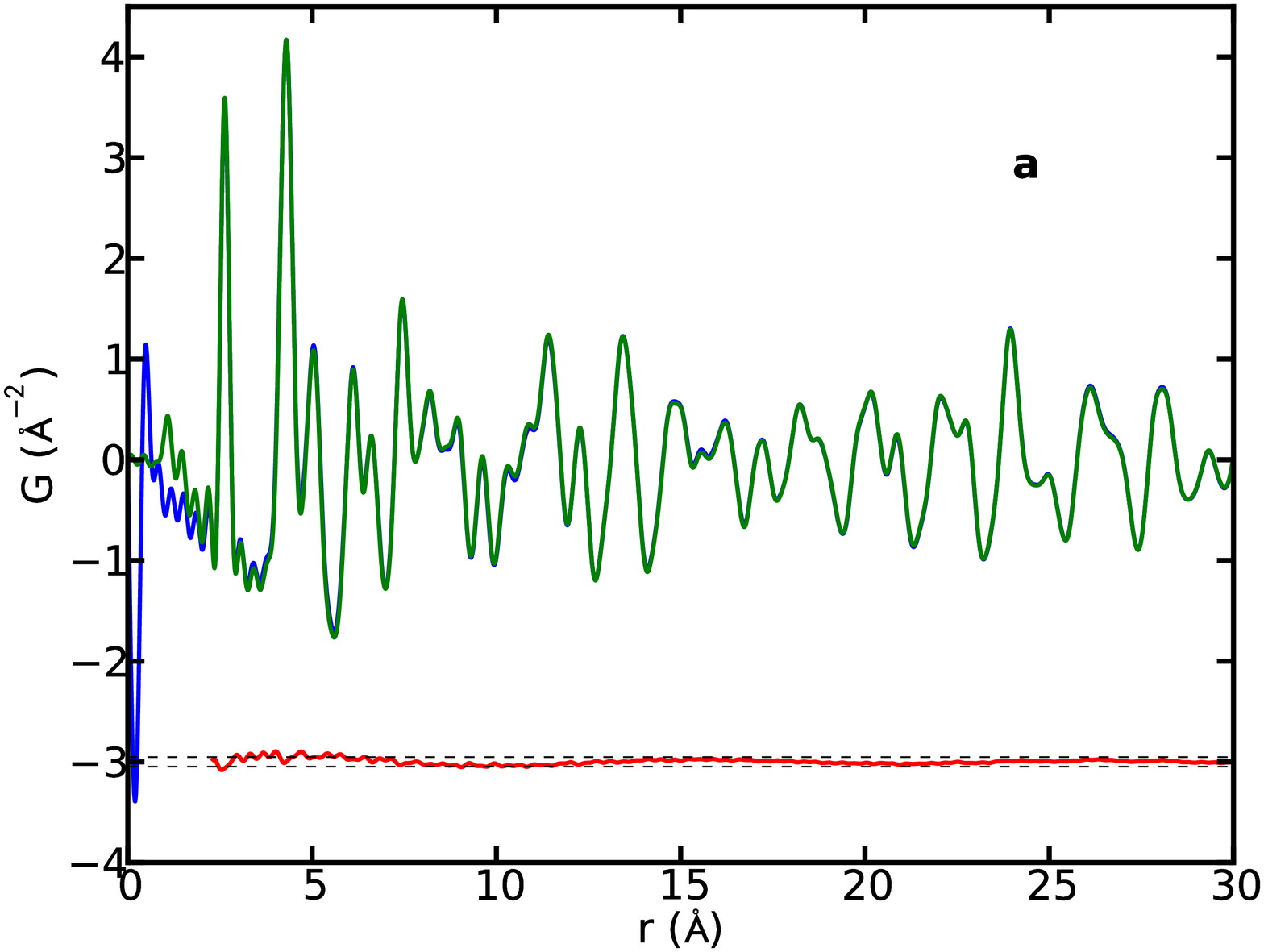}
        \includegraphics[width=0.5\textwidth]{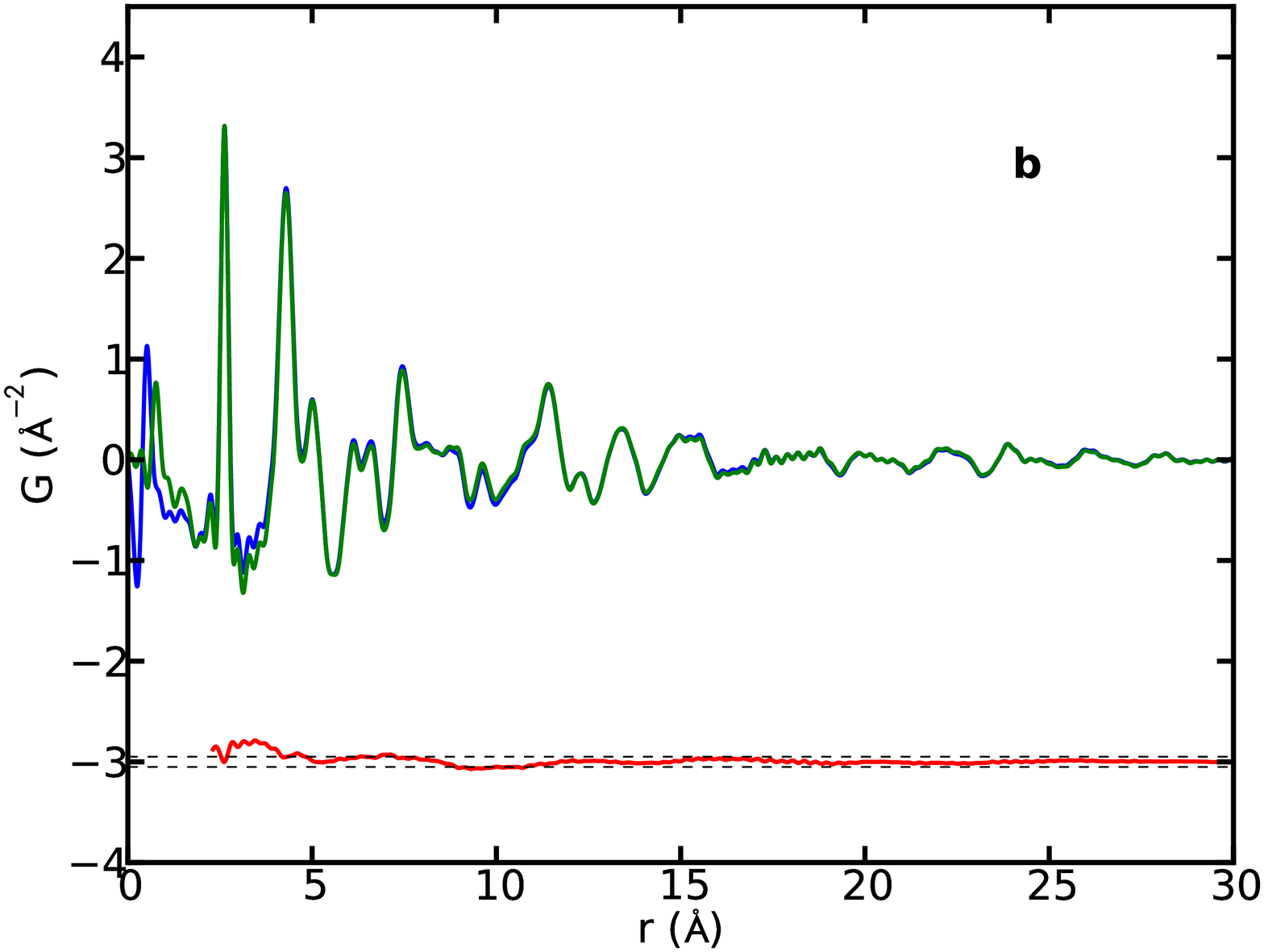}
        \includegraphics[width=0.5\textwidth]{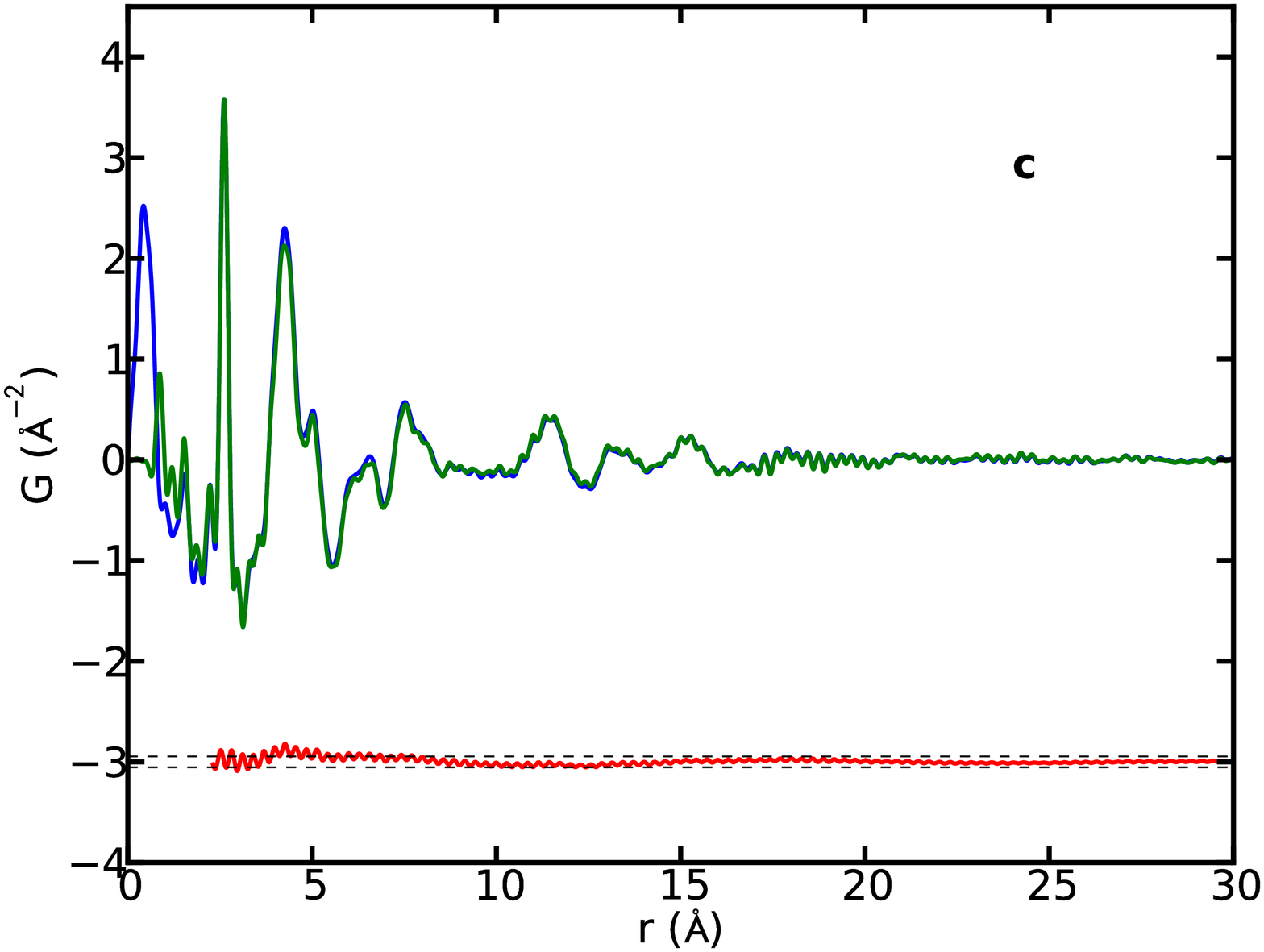}
    \caption{PDFs of \textbf{(a)} bulk CdSe and \textbf{(b)} 37~\AA, and \textbf{(c)} 22~\AA\ CdSe nanoparticles made with PDFgetX2 (blue) and PDFgetX3 (green) with $Q_{max}$ = 18.0~\AA$^{-1}$ in all cases. Difference curve (offset) is in red. The dashed lines represent two standard deviations in the difference curve ($r$ values below the nearest neighbor peaks were not included in the standard deviation calculation and, for the nanoparticle in panel \textbf{(c)}, $r$ values larger than 22~\AA\ were not included).}
    \label{fig;cdse-compare}
\end{figure}

We now turn our attention to a more challenging class of materials: nanoparticles which tend to be weakly scattering and more disordered.
In Figure~\ref{fig;cdse-compare} we show PDFs of three samples of cadmium selenide (CdSe) taken from data  published in \cite{masad;prb07}. The bulk CdSe in panel \textbf{(a)} is included for completeness. The nanoparticles in panels \textbf{(b)} and \textbf{(c)} were calculated to have diameters of 37~\AA\ and 22~\AA, respectively \cite{masad;prb07}.
We see that in all three panels of Figure~\ref{fig;cdse-compare}, the PDFs made with the two programs are almost identical. It was challenging to obtain the PDFs from PDFgetX2 requiring considerable care and user intervention and parameter tuning.  In the case of PDFgetX3 the PDFs shown were produced with no more effort than the Ni and BaTiO$_3$ PDFs shown above. The low-$r$ region looks a little bit different between the PDFs, especially as the size of the nanoparticles gets smaller, but we remember that this region contains no physical information. In fact, we might even argue that in panel \textbf{(c)}, the PDFgetX3 PDF looks cleaner than the PDFgetX2 PDF.

Table~\ref{tab;cdse-fit} contains the refined parameters for the CdSe samples compared to a model based on wurtzite \cite{wycko;bk67}. Again we see very good agreement between all parameters determined from the getX2 and PDFgetX3 PDFs and the residual, $R_w$, is highly comparable between the two pairs of PDFs.

\begin{table}
\centering
\caption{Comparison of the parameters refined in fitting the CdSe wurtzite model to the PDFs.}
\label{tab;cdse-fit}
\begin{tabular}{l . .}
Parameter  &  \dothead{PDFgetX2}  &  \dothead{PDFgetX3}  \\
\hline
Bulk && \\
\hline
\Qdamp\ (\AA$^{-1}$)                    & 0.0593    & 0.0599 \\
$a=b$ (\AA)                             & 4.2996    & 4.2996 \\
$c$ (\AA)                               & 7.0112    & 7.0113 \\
$\delta_{2}$ (\AA$^{2}$)                & 3.21      & 3.26   \\
$U_{11,\mathrm{Cd}}$ = $U_{22,\mathrm{Cd}}$ (\AA$^{2}$)
                                        & 0.0156    & 0.0155 \\
$U_{33,\mathrm{Cd}}$ (\AA$^{2}$)        & 0.0143    & 0.0141 \\
$U_{11,\mathrm{Se}}$ = $U_{22,\mathrm{Se}}$ (\AA$^{2}$)
                                        & 0.0129    & 0.0128 \\
$U_{33,\mathrm{Se}}$ (\AA$^{2}$)        & 0.0581    & 0.0575 \\
$R_w$                                   & 0.114     & 0.104  \\
\hline
37~\AA\ nanoparticle && \\
\hline
$a=b$ (\AA)                             & 4.2956    & 4.2961 \\
$c$ (\AA)                               & 7.0068    & 7.0075 \\
$\delta_{2}$ (\AA$^{2}$)                & 4.66      & 4.74   \\
$U_{11,\mathrm{Cd}}$ = $U_{22,\mathrm{Cd}}$ (\AA$^{2}$)
                                        & 0.0225    & 0.0221 \\
$U_{33,\mathrm{Cd}}$ (\AA$^{2}$)        & 0.0302    & 0.0302 \\
$U_{11,\mathrm{Se}}$ = $U_{22,\mathrm{Se}}$ (\AA$^{2}$)
                                        & 0.0120    & 0.0118 \\
$U_{33,\mathrm{Se}}$ (\AA$^{2}$)        & 0.199     & 0.194  \\
Particle diameter (\AA)                 & 36.39     & 35.34  \\
$R_w$                                   & 0.194     & 0.173  \\
\hline
22~\AA\ nanoparticle && \\
\hline
$a=b$ (\AA)                             & 4.2940    & 4.2948 \\
$c$ (\AA)                               & 6.8567    & 6.8633 \\
$\delta_{2}$ (\AA$^{2}$)                & 4.97      & 5.20   \\
$U_{11,\mathrm{Cd}}$ = $U_{22,\mathrm{Cd}}$ (\AA$^{2}$)
                                        & 0.0433    & 0.0415 \\
$U_{33,\mathrm{Cd}}$ (\AA$^{2}$)        & 0.0403    & 0.0409 \\
$U_{11,\mathrm{Se}}$ = $U_{22,\mathrm{Se}}$ (\AA$^{2}$)
                                        & 0.0199    & 0.0203 \\
$U_{33,\mathrm{Se}}$ (\AA$^{2}$)        & 0.233     & 0.221  \\
Particle diameter (\AA)                 & 23.13     & 23.35  \\
$R_w$                                   & 0.262     & 0.265  \\
\hline
\end{tabular}
\end{table}

\subsection{Pharmaceuticals}

\begin{figure}
    \centering
        \includegraphics[width=0.6\textwidth]{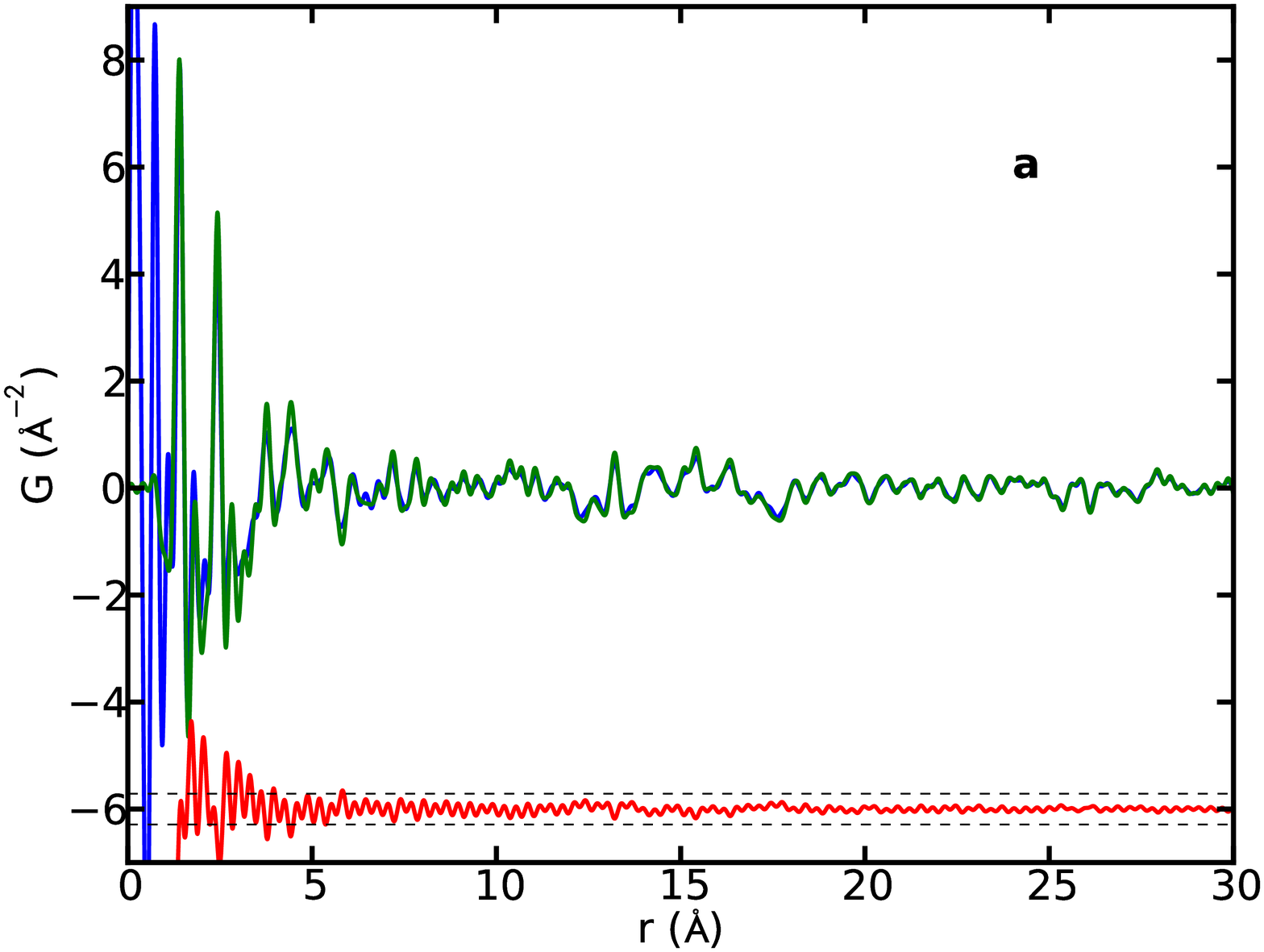}
        \includegraphics[width=0.6\textwidth]{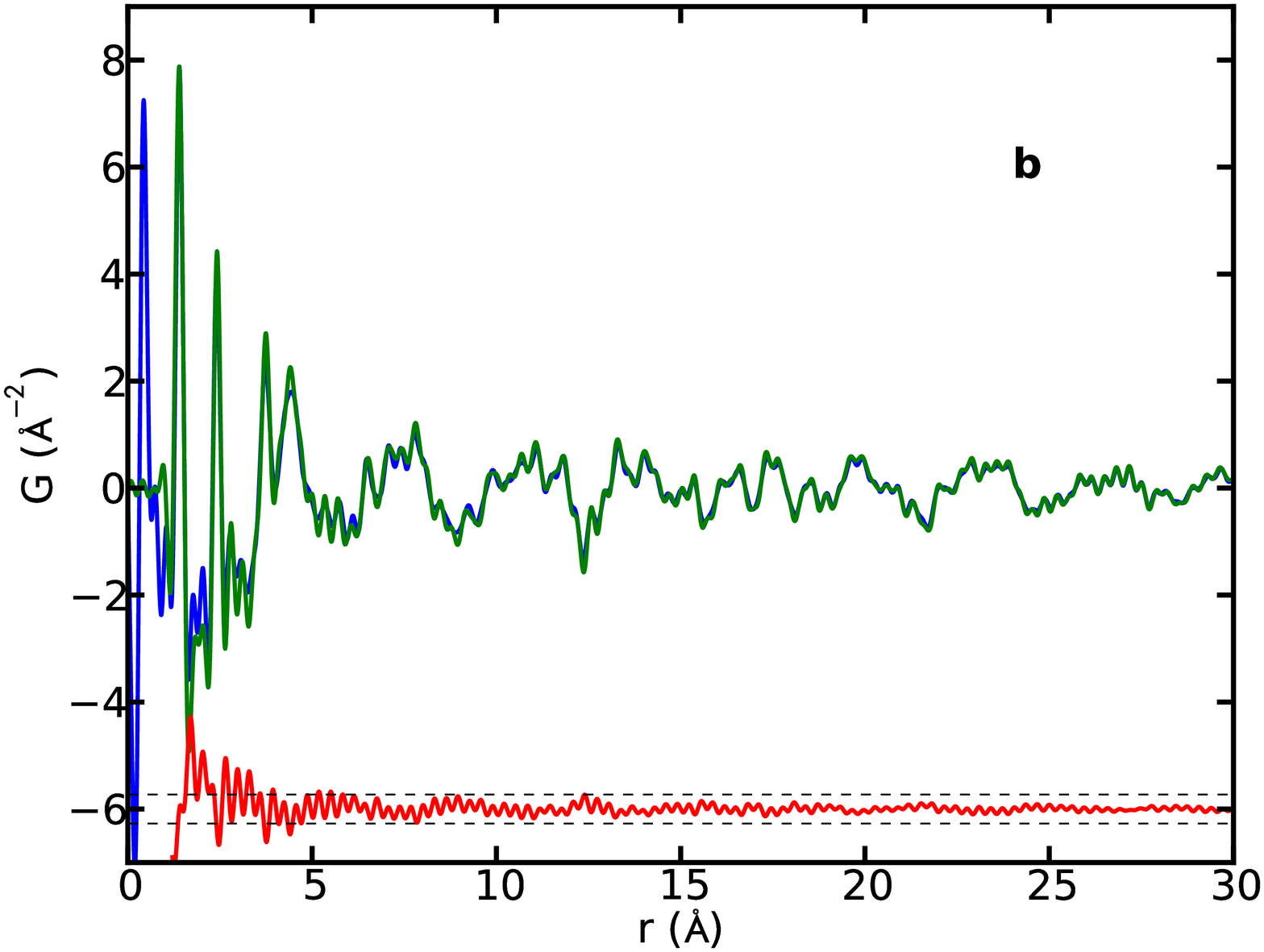}
        \includegraphics[width=0.6\textwidth]{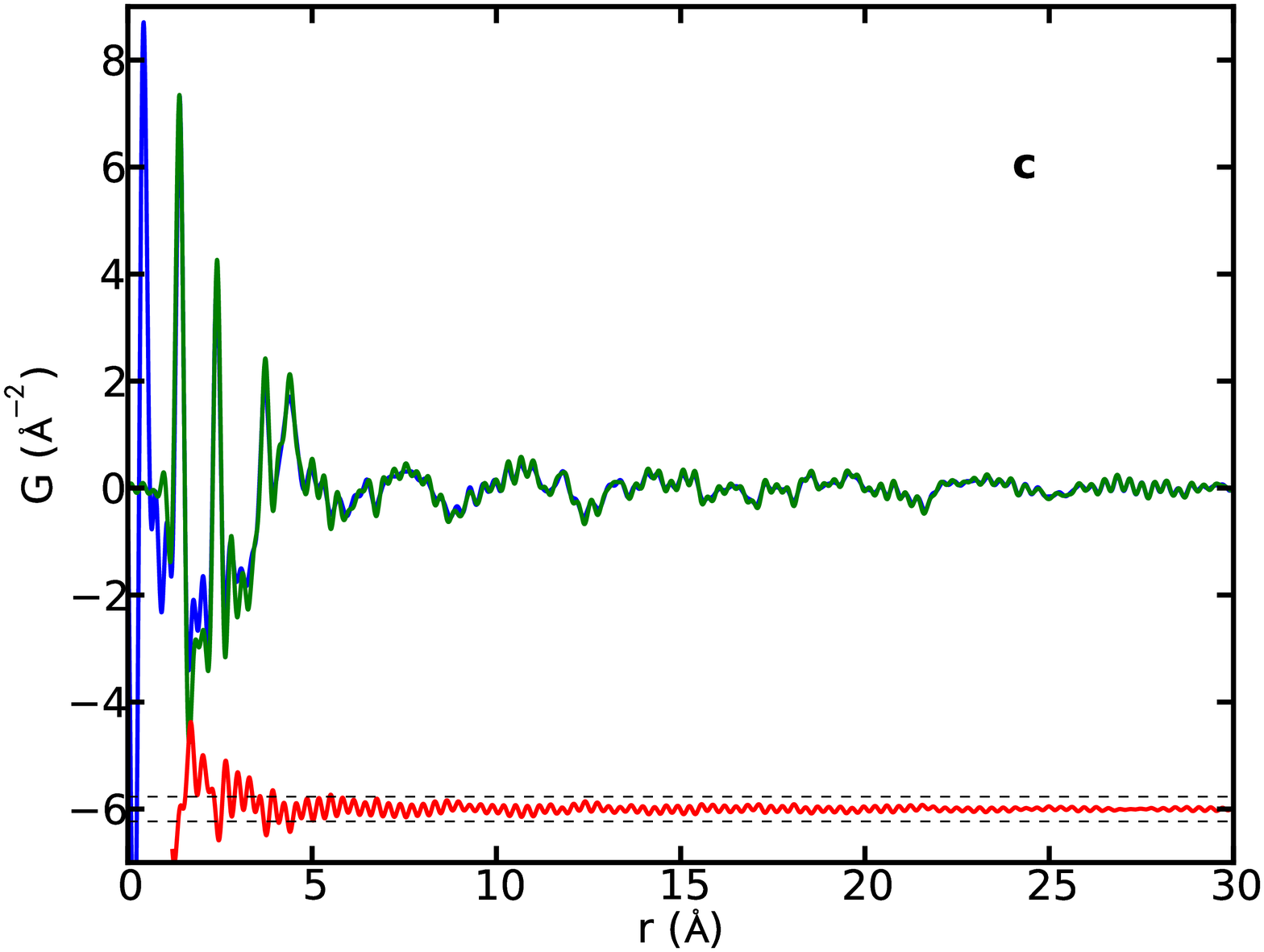}
    \caption{PDFs of \textbf{(a)} CBZ-I, \textbf{(b)} CBZ-III, and \textbf{(c)} nanostructured CBZ made with PDFgetX2 (blue) and PDFgetX3 (green) with $Q_{max}$ = 20.0~\AA$^{-1}$ in all cases. Difference curve (offset) is in red. The dashed lines represent two standard deviations in the difference curve. ($r$ values below the nearest neighbor peaks were not included in the standard deviation calculation).}
    \label{fig;drugs-compare}
\end{figure}

The final class of materials that we tested are organic pharmaceutical compounds. These materials can be crystalline, as we see in Figure~\ref{fig;drugs-compare}\textbf{(a)} and \textbf{(b)}, nanostructured as in Figure~\ref{fig;drugs-compare}\textbf{(c)}, or amorphous. These materials tend to have relatively complicated crystal structures that are made up of mostly light, organic elements such as hydrogen, carbon, and oxygen that do not diffract strongly and even crystal phase pharmaceutical compounds require quite a bit of tinkering in PDFgetX2 to produce a good PDF.

In the examples here we consider three polymorphs of the drug carbamazepine (CBZ), crystalline CBZ form-I and form-III as well as melt-quenched carbamazepine that turned out to be nanocrystalline \cite{billi;cec10,dykhn;phmr11}.
As with the nanoparticles in Figure~\ref{fig;cdse-compare}, the PDFs in Figure~\ref{fig;drugs-compare} made with PDFgetX2 have relatively large fluctuations from imperfect corrections at low $r$. This is common for weakly scattering samples.  However, these were the best PDFs that could be obtained using PDFgetX2 at the time of publication. The PDFs made with PDFgetX3 are highly similar in the physically meaningful region above the nearest neighbor separation (C-C bond at 1.4 \AA) with the added benefit that they appear to be cleaner in the unphysical low-r region.  This is an advantage because termination ripples from large features in the unphysical region may propagate into the physically meaningful region of the PDF.

We did not fit these PDFs to models because new modeling tools need to be developed for this class of materials.

\section{Summary}
We have described and demonstrated an implementation of the {\it ad-hoc} data reduction protocol described in~\cite{billi;aca12} in  a new Python based software program PDFgetX3.  PDFs obtained using this method have been compared with PDFs obtained using PDFgetX2, an established program for producing PDFs, and are found to be highly similar.  Models fit to the PDFgetX2 and PDFgetX3 PDFs yield refined parameters that are correspondingly similar.  The program has been tested on a range of samples from strongly scattering inorganic crystalline powders such as nickel and BaTiO$_3$ to weakly scattering low atomic number pharmaceutical compounds.  The program is easy to use compared to PDFgetX2 and rapid, giving PDFs in real-time as parameters such as background scale or $Q_{max}$ are varied.  The program should be good for most PDF studies (though does not yield data on an absolute scale), but will prove to be especially useful for high throughput studies such as parametric or time-resolved experiments.   More information about the program is available at www.diffpy.org.


\end{document}